\documentclass[preprint,aps,amsmath,amssymb,12pt,showpacs,showkeys]{revtex4} 

\usepackage{bbm}
\usepackage[dvips]{graphicx}
\usepackage{color}
\usepackage{amsmath,amssymb,latexsym}

\newcommand{\micron}{\ensuremath{\mu\mathrm{m}}}

\newcommand{\e}{\mathbf{e}}
\newcommand{\f}{\mathbf{f}}
\newcommand{\rr}{\mathbf{r}}
\newcommand{\vv}{\mathbf{v}}

\newcommand{\bigO}{\mathcal{O}}

\newcommand{\what}[1]{\widehat{#1}}

\newcommand{\D}{\mathcal{D}}
\newcommand{\E}{\mathcal{E}}
\newcommand{\F}{\mathcal{F}}
\newcommand{\G}{\mathcal{G}}
\newcommand{\R}{\mathcal{R}}

\begin{document}

\title{Flagellar swimmers oscillate between pusher- and puller-type swimming}

\author{Gary S. Klindt}
\affiliation{Max Planck Institute for the Physics of Complex Systems, Dresden, Germany}
\author{Benjamin M. Friedrich}
\email{benjamin.friedrich@pks.mpg.de}
\affiliation{Max Planck Institute for the Physics of Complex Systems, Dresden, Germany}

\date{\today}



\begin{abstract}
Self-propulsion of cellular microswimmers generates flow signatures, 
commonly classified as pusher- and puller-type, which characterize hydrodynamic interactions with other cells or boundaries. 
Using experimentally measured beat patterns, we compute that flagellated alga and sperm oscillate between pusher and puller. 
Beyond a typical distance of $100\,\micron$ from the swimmer, 
inertia attenuates oscillatory micro-flows.
We show that hydrodynamic interactions between swimmers oscillate in time and are of similar magnitude as stochastic swimming fluctuations. 
\end{abstract}

\maketitle


For a single cell swimming in a fluid, inertia is negligible
\cite{lauga_hydrodynamics_2009,elgeti_physics_2014}.
Cellular swimmers exploit propulsion strategies independent of inertia
that allow for net propulsion by viscous forces~\cite{geoffrey_taylor_analysis_1951-1}.
Periodic, non-reciprocal body shape changes~\cite{howard_c._berg_bacteria_1973, shaevitz_spiroplasma_2005, gray_propulsion_1955, ruffer_high-speed_1985}
conquer the time-reversibility of the Stokes equation,
which governs fluid flow in the inertia-less limit of zero Reynolds number. 
For example,
traveling bending waves of long slender cell appendages termed flagella
propel many eukaryotic cells, 
including sperm or swimming alga~\cite{brennen_fluid_1977}.
Self-propulsion by such periodic shape changes implies that 
net motion is a second-order-effect,
superimposed to an oscillatory motion, 
first-order in the amplitude of the swimming stroke~\cite{shapere_self-propulsion_1987,polotzek_three-sphere_2013}.

\begin{figure}[ht]
    \begin{center}
        \includegraphics[width=0.5\textwidth]{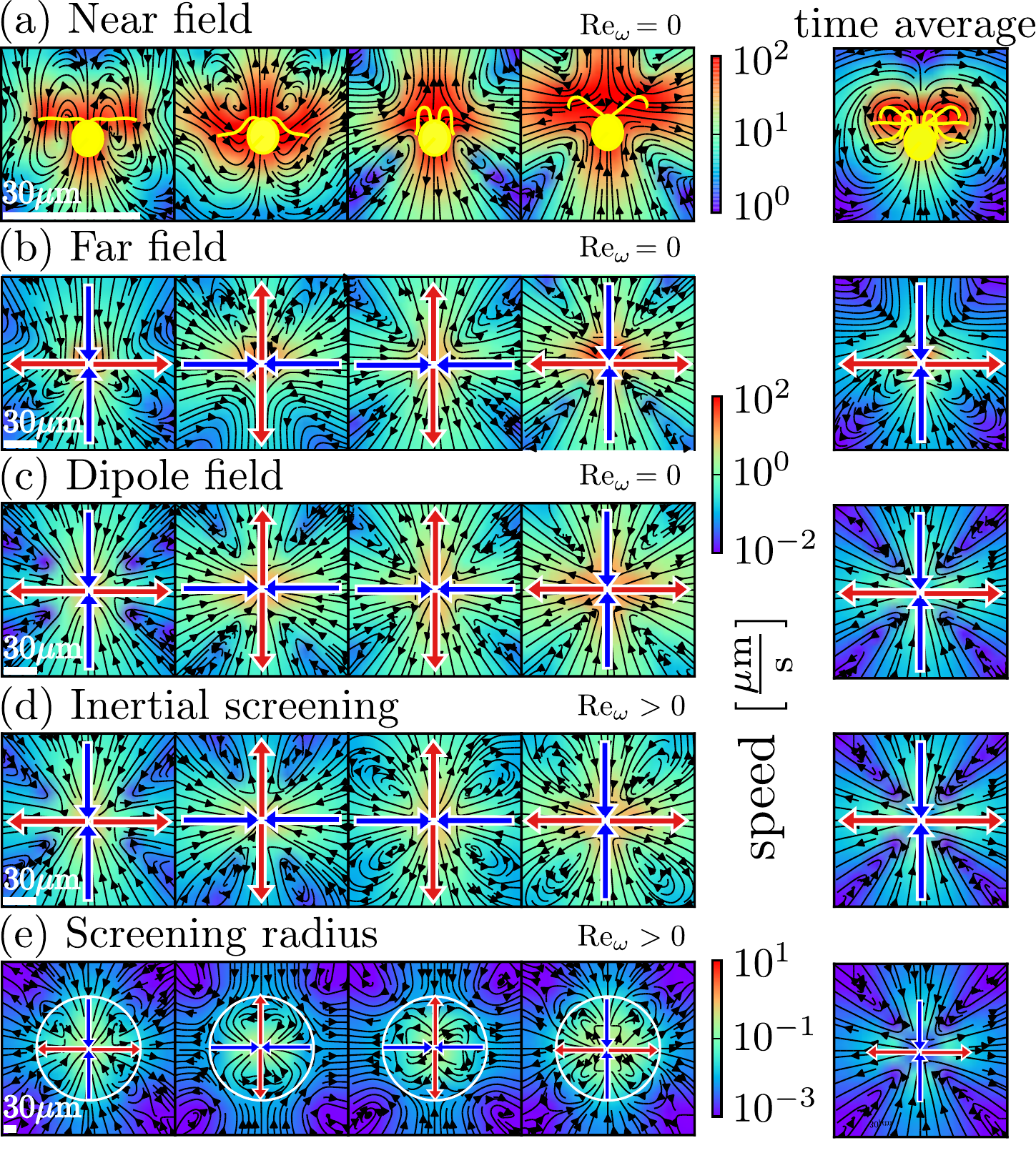}
    \end{center}
\caption[]{
\label{fig:flowfields}
Time-dependent fluid flows induced by a \textit{Chlamydomonas} cell swimming in water.
\textbf{(a,b)}
Near field and far field of fluid flow, computed in the limit of zero Reynolds number, 
corresponding to the idealization of an inertia-less fluid. 
\textbf{(c)}
The far field is faithfully reproduced by its leading order singularity, 
the Stokes doublet flow induced by a localized force dipole,
cf.~Eq.~(\ref{eq:multipoleFlow}).
\textbf{(d,e)}
Beyond a characteristic distance $\delta$ from the swimmer, 
inertial effects attenuate oscillatory micro-flows,
resulting in a deviation from the Stokes doublet flow.
Shown are solutions of the linearized Navier-Stokes equation,  
Eq.~(\ref{eq:inertial_flow}),
induced by the oscillating force dipole from panel (c).
The white circle in (e) has radius $3\delta$.
Beat frequency: $\omega/(2\pi)=50\,\mathrm{Hz}$.
}
\end{figure}

The flow fields generated by microswimmers 
are commonly classified as pusher- or puller-type,
depending on the direction of fluid flow along
the axis of net swimming~\cite{downton_simulation_2009,spagnolie_hydrodynamics_2012,li_hydrodynamic_2014}:
inward flow towards the swimmer characterizes a puller,
while outward flow characterizes a pusher, 
see Fig.~\ref{fig:flowfields}(b).
This characterization allows a simple assessment
of the interaction of microswimmers with boundary surfaces
or inter-swimmer interactions. 
For example, it was suggested that pushers become hydrodynamically 
attracted to boundary surfaces by an inward flow perpendicular to the axis of net swimming~\cite{lauga_hydrodynamics_2009}.
Swimmer type further determines active rheological responses 
such as shear thinning in dense suspensions of micro\-swimmers~\cite{rafai_effective_2010,gluzman_effective_2013}.

Here, we compute time-varying flow fields induced by flagellated swimmers, 
using experimentally measured beat patterns~\cite{geyer_cell-body_2013,friedrich_high-precision_2010}.
We show that flagellated micro\-swimmers periodically oscillate between pusher- and puller-type swimming,
which implies more complex, dynamic interactions with boundaries and other swimmers.
Further, we discuss how inertial effects, usually neglected in microswimming problems,
attenuate oscillatory micro-flows at a distance $\delta\sim 100\,\micron$ from the swimmer.
For that aim, we present a novel approximation scheme to account for inertial effects in microswimming.
From our analysis, we find that not only do flow fields oscillate, 
but also the rate of hydrodynamic dissipation associated with flagellar swimming.
This simple fact has important consequences~\cite{alouges_optimal_2009}. 
It implies that flagellar beat patterns are not optimized to minimize hydrodynamic
dissipation as sole optimization criterion,
as considered previously~\cite{guo_cilia_2014, spagnolie_optimal_2010, pironneau_optimal_1974}.

\paragraph{Time-dependent micro-flows.} 
Swimming by periodic body shape changes results in time-varying flow fields, 
which have been measured experimentally
for a flagellated model swimmer, 
the bi-flagellated green alga \textit{Chlamydomonas} \cite{guasto_oscillatory_2010}.
Using experimentally measured beat patterns~\cite{geyer_cell-body_2013}, 
we computed time-dependent flow fields, 
first for the limit of zero Reynolds number, see Figure~\ref{fig:flowfields}(a,b).
Specifically, we used a fast boundary element method~\cite{liu_fast_2006} to determine
the surface density of forces $\f(\rr,t)$ exerted by the time-dependent surface $S(t)$ of the swimmer.
The dynamics of $S(t)$ is a superposition of an imposed shape change of the two flagella, 
and a rigid body motion of the whole swimmer, which is solved self-consistently
to ensure force and torque balance of the whole cell.

In the limit of zero Reynolds number, 
the fluid flow field $\vv(\rr,t)$
is obtained by propagating $\f(\rr,t)$
with the Green's function $\mathbf{G}$ of viscous flow, 
$v_i(\rr,t) = \int_{S(t)}\! d^2\rr'\, G_{ij}(\rr-\rr') f_j(\rr',t)$.
Here,
$G_{ij}(\rr)=(\delta_{ij}+\what{\rr}_i\what{\rr}_j)/(8\pi\eta r)$,
is known as the Oseen tensor,
$\eta$ is fluid viscosity, 
and $r=|\rr|$, $\what{\rr}=\rr/r$.
Next, we show how this flow field can be decomposed into fundamental singularities
with an oscillating force dipole characterizing the far field.

\paragraph{Hydrodynamic multipole expansion.}
The force density $f_j(\rr)$ can be decomposed into Cartesian multipoles 
$\F_{j,J}(t) = \int_{S(t)} \! d^2\rr\, (\rr-\rr_0)^J f_j(\rr, t)$ \cite{ghose_irreducible_2014},
similar to the decomposition of a charge distribution into charge multipoles.
Here, $J=(j_1,\ldots,j_k)$ denotes a multi-index and we use standard multi-index notation.
The force density is formally recovered from its multipoles as
$f_j(\rr) = \sum_J (-1)^{|J|} \F_{j,J} \nabla^J \delta(\rr-\rr_0)/J!$~\cite{schmitz_force_1980}.
We can thus represent the flow field as a superposition of fundamental singularities
\begin{align}
\label{eq:multipoleFlow}
v_i(\rr) = \sum_J \frac{(-1)^{|J|}}{J!} \nabla^J G_{ij} (\rr-\rr_0) \F_{j,J}.
\end{align}
These singularities are traditionally known as Stokeslet ($|J|=1$), Stokes doublet ($|J|=2$)~\cite{ghose_irreducible_2014}.
Note that a self-propelled swimmer, being free from external forces and torques, 
does not exert any net force or torque on the surrounding fluid. 
Thus, the force monopole $\F_j$ vanishes, 
and the force dipole matrix $\F_{j,k}$ is symmetric.
This force dipole represents the leading order singularity, which characterizes the far field of fluid flow, 
decaying as $r^{-2}$ with distance $r$ from the swimmer.
For an incompressible fluid, the trace of $\F_{j,k}$ does not contribute to the flow field, but describes a localized pressure.
Without loss of generality, we assume that $\F_{j,k}$ is traceless.

The force multipoles change in time, as the microswimmer changes shape and position.
The leading order singularity $\F_{j,k}$ 
oscillates with the beat frequency $\omega$ 
\endnote{For force-free swimmers, $\F_{j,k}$ is independent of $\rr_0$.}.

We find that higher order multipoles dominate the near field on length-scales comparable to the size of the swimmer,
while the dipole contribution faithfully reproduces the far field 
at distances of $10-100\,\micron$, see Fig.~\ref{fig:flowfields}(a-c).
At even larger distances, inertial effects cannot be neglected anymore, 
as we discuss next.

\paragraph{Screening of oscillatory micro-flows by fluid inertia.}
The full nonlinear Navier-Stokes equation of fluid flow contains both viscous and inertial terms
$\rho\dot{\vv}+\rho\vv\cdot\nabla\vv = - \nabla p + \eta\Delta\vv$.
The inertial terms on the left-hand-side are usually negligible in the proximity of a microswimmer.
Their relative magnitude compared to viscous terms is approximated by the
instationary Reynolds number, $\mathrm{Re}_\omega=\rho\omega A^2/\eta$.
For a typical flagellated swimmer in water, $\mathrm{Re}_\omega$ is small:
$\mathrm{Re}_\omega=10^{-2}$ 
for amplitude $A=5\,\micron$, 
beat frequency $\omega/(2\pi)=50\,\mathrm{s}^{-1}$, 
fluid density $\rho=10^3\,\mathrm{kg}/\mathrm{m}^3$, 
viscosity $\eta=10^{-3}\,\mathrm{Pa}{\cdot}\mathrm{s}$.
Yet, inertial effects become important at a distance.
For the flow field induced by a force dipole, 
the unsteady acceleration
$\rho\dot{\vv}\sim\rho\omega r^{-2}$
decays slower than the viscous forces 
$\eta\Delta\vv\sim r^{-4}$,
while the convective acceleration decays like
$\rho \vv\cdot\nabla\vv\sim r^{-5}$.
Inertial effects due to unsteady acceleration are thus expected
to become important beyond a distance $\delta=[2\eta/(\rho\omega)]^{1/2}$ \cite{landau_fluid_1987}.
Using the above values, 
we estimate $\delta\sim 100\,\micron$.

This argument further shows that convective acceleration can be neglected throughout the fluid.
Hence, the linearized Navier-Stokes equation applies
\begin{equation}
    \label{eq:lin_navier_stokes}
    \rho\dot{\vv} = -\nabla p + \eta \Delta \vv.
\end{equation}
The flow induced by an oscillating force had been studied by Stokes more than a century
ago~\cite{stokes_effect_1851, landau_fluid_1987}.
In modern language, an oscillating force monopole $f_j \exp(-i\omega t)$
induces an oscillating flow field 
$v_i(t)(\rr)=\G_{ij}(\rr) f_j\exp(-i\omega t)$ 
characterized by the instationary Stokes tensor
\begin{equation}
    \label{eq:inst_stokes}
    \G_{ij}(\rr,\omega) = \frac{\E}{8\pi\eta r} 
    \left[
        (1-i\D)\delta_{ij} + (1+3i\D) \hat{r}_i \hat{r}_j 
    \right],
\end{equation}
where
$\E = \exp[-(1-i)r/\delta]$ and
$\D = (\delta/r)^2 [\E^{-1} - 1 - (1-i)(r/\delta) - (1-i)^2(r/\delta)^2/2 ]$.
For a Fourier transform, see \cite{espanol_oscilet_1995}.
The near field with $r\ll\delta$,
reproduces the Oseen tensor
$v_i(\rr) = G_{ij}(\rr)f_j\exp(-i\omega t) + \bigO(r/\delta)$.
For the far field with $r\gg\delta$,
we obtain a potential flow
that decays as $\sim r^{-3}$ with
$v_i=-3i\delta \Delta G_{ij}(\rr)f_j\exp(-i\omega t)+\bigO(\delta/r)^4$.
The rotational part of the Stokeslet flow field is exponentially attenuated
as $\exp(-r/\delta)$.
In analogy to the common name `Stokeslet' for flow fields of the form $G_{ij}f_j$,
one may call the oscillatory flow field $\G_{ij}f_j\exp(-i\omega t)$ `oscilet'.

In the limit $\delta\gg L$, 
we can estimate the oscillating force dipole $\F_{j,k}$ 
exerted by the microswimmer using the limit of zero Reynolds number as employed above.
The far field according to Eq.~(\ref{eq:lin_navier_stokes}), 
including inertial effects due to unsteady acceleration, 
is then obtained as a superposition of oscilet derivatives,
\begin{equation}
    \label{eq:inertial_flow}
    v_i^\mathrm{far}(\rr) 
    = \sum_n \partial_k \G_{ij}(\rr,n\omega) \widetilde{\F}_{j,k}^{(n)} \exp(-in\omega t),
\end{equation}
where 
$\F_{j,k}=\sum_n \widetilde{\F}_{j,k}^{(n)} \exp(-i\omega n t)$.
Figure~\ref{fig:flowfields}(d,e) show resultant flow fields for 
\textit{Chlamdyomonas}, exemplifying the effect of inertial screening.

\paragraph{Oscillating between pusher and puller.}
We introduce a coordinate system $(\e_1,\e_2,\e_3)$ of ortho-normal vectors, 
slowly co-moving with the microswimmer,
such that $\e_1$ points along the net swimming direction.
With this choice of coordinates, 
the component $\F_{11}$ of the force dipole tensor determines 
the flow speed along the swimming direction
as $v_1=3\F_{11}/(8\pi\eta r^2)$ (in the limit of zero Reynolds number, neglecting higher multipoles).
Thus, the force dipole tensor allows to 
discriminate pusher-type ($\F_{11}>0$) and puller-type ($\F_{11}<0$) swimmers.
More generally, 
we can write the traceless, symmetric force dipole tensor as
\begin{equation}
\left\{
\F_{j,k}
\right\} 
    = f_s\, \mathfrak{R}
    \left[
        \left(
        \begin{matrix}
            2 & 0 & 0 \\
            0 & -1 & 0 \\
            0 & 0 & -1
        \end{matrix}
        \right)
        +
        \left(
        \begin{matrix}
            0 & 0 & 0 \\
            0 & \epsilon & 0 \\
            0 & 0 & -\epsilon
        \end{matrix}
        \right)
    \right]       
	\mathfrak{R}^T,
    \label{eq:tracelessdipoledecomposition3d}
\end{equation}
where 
$f_s$ denotes a scalar dipole strength, 
$\epsilon$ a dimensionless asymmetry parameter, 
and $\mathfrak{R}$ a rotation matrix.

\begin{figure}[ht]
        \includegraphics[width=0.5\textwidth]{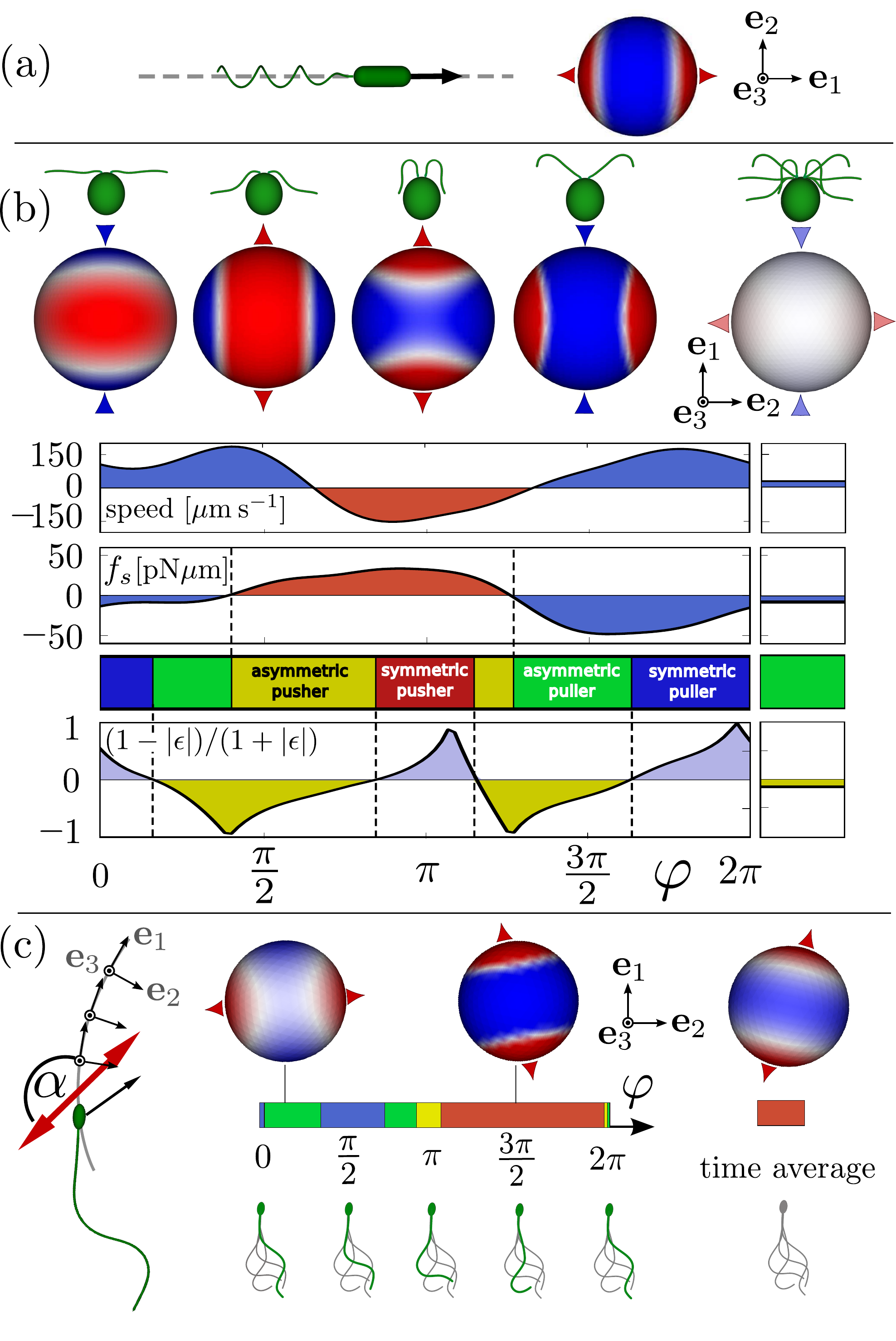}
\caption[]{
\textbf{(a)}
	A bacterium propelled by a rotating helical filament
    represents a symmetric pusher.    
\textbf{(b)}
        \textit{Chlamydomonas} oscillates between puller- and pusher-type swimming.
        From top to bottom (fifth column shows time-averages):
        Flagellar shapes at equidistant phases of the beat cycle.       		  
        Radial projections of the induced Stokes doublet flow for each of these phases, 
		cf.~Eq.~(\ref{eq:multipoleFlow}).
        Instantaneous swimming velocity as a function of flagellar phase. 
        Dipole strength $f_s$, cf.~Eq.~(\ref{eq:tracelessdipoledecomposition3d}).
        Flow signatures are classified as either 
		symmetric or asymmetric pusher- or puller-type swimmers, 
        depending on whether the rescaled asymmetry parameter 
        $(1-|\epsilon|)/(1+|\epsilon|)$ is positive or negative. 
\textbf{(c)}
    A swimming sperm oscillates between pusher and puller, 
    with a force dipole whose orientation rotates during the beat cycle.
    }
    \label{fig:chlamyDipole}
\end{figure}

For a microswimmer with special symmetries, 
$\mathfrak{R}$ can be chosen particularly simple. 
The idealized bacterium shown in figure~\ref{fig:chlamyDipole}(a)
has rotational symmetry around the $\e_1$-axis,
and we may choose $\mathfrak{R}=\mathbb{I}$ and $\epsilon=0$.
We find $f_s>0$, 
confirming this swimmer as a pusher.

Next, for a swimmer with double mirror-symmetry,
such as \textit{Chlamydomonas} with symmetric breast-stroke beat, 
we can choose $\mathfrak{R}{=}\mathbb{I}$.
In this case, the sign of $f_s$ determines pusher-type ($f_s>0$) or puller-type ($f_s<0$).
Remarkably, we find that the sign of $f_s$ alternates during a beat cycle,
being negative during the effective stroke, 
but positive during the recovery stroke, see Fig.~\ref{fig:chlamyDipole}(b).
Further, the asymmetry parameter $\epsilon$ is found to transiently exceed unity, 
implying that the flow in the $(\e_2,\e_3)$-plane normal to the swimming direction
changes sign as a function of azimuthal angle.
We will use the terms
`asymmetric pusher' and `asymmetric puller' for a microswimmer with $|\epsilon|>1$. 
Note that a smooth transition from a pusher to a puller generally passes 
through an asymmetric swimmer-type if $\mathfrak{R}{=}\mathbb{I}$.
The time-averaged force dipole 
exerted by \emph{Chlamydomonas} corresponds to an asymmetric puller.

Finally, for a swimmer with single mirror-symmetry, 
say with respect to a plane with normal $\e_3$, 
we can choose $\mathfrak{R}$ as a rotation around the $\e_3$ axis by an angle $\alpha$.
This case applies to sperm cells with planar beat patterns \cite{friedrich_high-precision_2010}, 
swimming in the $(\e_1,\e_2)$-plane.
Now, $\F_{11}=f_s [2\cos^2\alpha+(\epsilon-1)\sin^2\alpha]$.
Figure~\ref{fig:chlamyDipole}(c) shows the time-dependent force dipole exerted by a sperm cell, 
revealing again an oscillation between pusher- and puller-type swimming. 

\begin{figure}
    \begin{center}
        \includegraphics[width=0.4\textwidth]{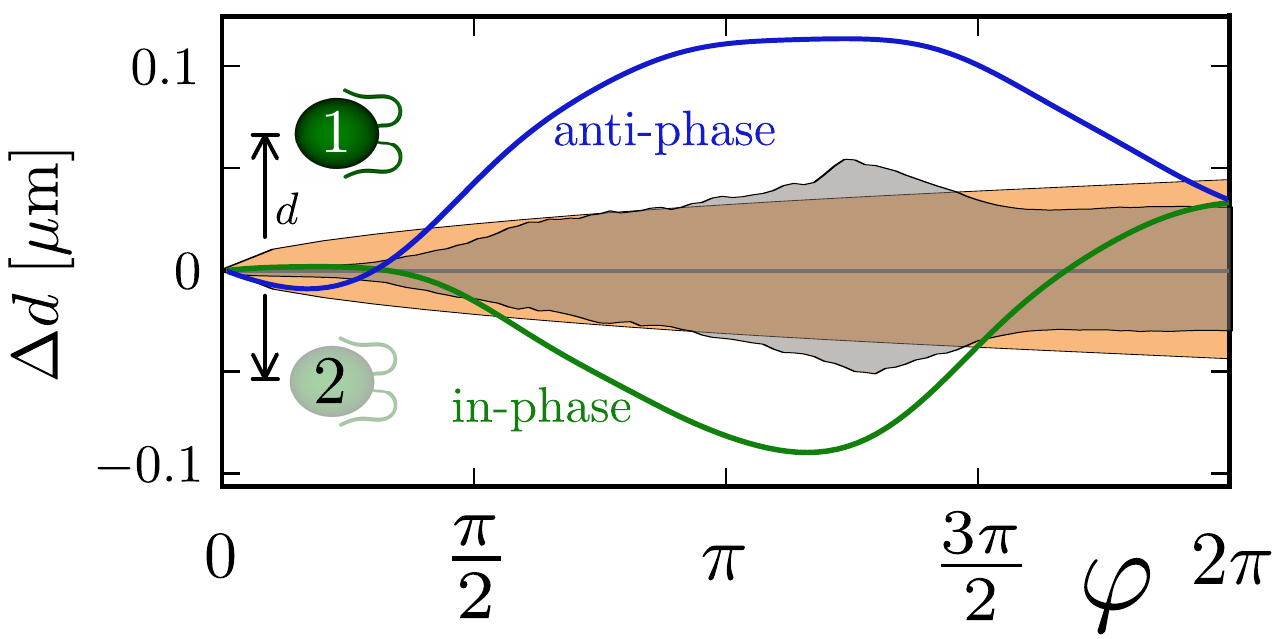}
    \end{center}
\caption[]{
Hydrodynamic interactions between microswimmers are weak and oscillatory.
Sidewards motion of a \textit{Chlamydomonas} cell
induced by hydrodynamic interactions with a second cell
that initially swims parallel at a distance $d = 24\,\micron$,
with swimming stroke either in-phase or anti-phase, respectively. 
For comparison, 
we show the expected {r.m.s.}\ side-wards displacement 
of a single cell due to passive thermal diffusion (orange shading),
or active amplitude fluctuations of the flagellar beat \cite{ma_active_2014} (gray).
}
    \label{fig:twoChlamies}
\end{figure}

\paragraph{Hydrodynamic interactions.}
Microswimmers can interact at a distance 
with other swimmers or a boundary wall by virtue of the flow they generate \cite{alexander_hydrodynamic_2010}.
It has been proposed that hydrodynamic interactions 
account for the mutual alignment of several swimmers,
beat synchronization in collections of flagella~\cite{yang_cooperation_2008},
or the active accumulation at boundary surfaces~\cite{smith_human_2009, elgeti_hydrodynamics_2010}.
The oscillatory nature of microflows generated by flagellated swimmers 
implies that these interactions likewise oscillate:
Fig.~\ref{fig:twoChlamies} shows periodic variations in the distance between 
two \textit{Chlamydomonas} cells swimming side-by-side.
These interactions strongly depend
on the relative phase difference of their flagellar beat.
While oscillatory flows dominate instantaneous interactions, 
their net contribution can be small compared to that induced by the static component of flow.
In fact, the interaction between two static dipoles separated by a distance $d$ scales as $d^{-2}$, 
while the net interaction between two oscillating dipoles 
scales as $d^{-5}$ after averaging over an oscillation cycle.
This is because the flow field induced by an oscillatory Stokes dipole (${\sim}d^{-2}$)
has to couple to the flow gradient (${\sim}d^{-3}$) to yield a net effect.
Further, hydrodynamic interactions 
can be masked by thermal and active noise, see Fig.~\ref{fig:twoChlamies}.

\paragraph{Phase-dependent rates of hydrodynamic dissipation.}
The viscous flow generated by a swimming cell
causes continuous dissipation of energy. 
For the flow field induced by an oscilet of strength $f_0\cos(\omega t)$, 
we find an oscillating rate of hydrodynamic dissipation
outside a spherical cut-off region of radius $R_0$ 
centered around the oscilet
\begin{equation}
\label{eq:Rosc}
    \mathcal{R}=\frac{f_0^2 \cos^2(\omega t)}{4\pi\eta}\left(
    \frac{1}{R_0} - \frac{1}{R_1}
    \right) 
    + \bigO\left(\frac{R_0}{\delta}\right)^2,
\end{equation}
where $R_1=(6/5)\delta$.
In fact, this value equals the viscous dissipation 
associated with a Stokeslet flow field induced by an oscillating force monopole $f_0\cos(\omega t)$
inside a spherical shell bounded by 
$R_0\le r\le R_1$.
Similar results are found for higher multipoles.
We conclude that the limit of zero Reynolds number
is appropriate for computing rates of hydrodynamic dissipation 
in the limit $L\ll\delta$.

Thus, neglecting inertial effects, 
the instantaneous power output of the swimming cell,
$\R=\int_S d^2\rr\,\f(\rr){\cdot}\vv(\rr)$,
equals the rate of hydrodynamic dissipation in the bulk of the fluid
$2\eta\int d^3\rr\, \nabla\vv(\rr){:}\nabla\vv(\rr)$.
Importantly, any swimming stroke that minimizes 
the mean power output $\langle{\R}\rangle=\oint d\varphi\,\R(\varphi)$
must satisfy $\R(\varphi){=}\mathrm{const}$~\cite{alouges_optimal_2009}.
Otherwise, reparameterizing
$\varphi'(\varphi)=
2\pi\int_0^\varphi d\theta\,\R(\theta)^{-1/2}/\int_0^{2\pi}d\theta\,\R(\theta)^{-1/2}$
reduces $\langle\R\rangle$. 

\begin{figure}[ht]
    \includegraphics[width=0.45\textwidth]{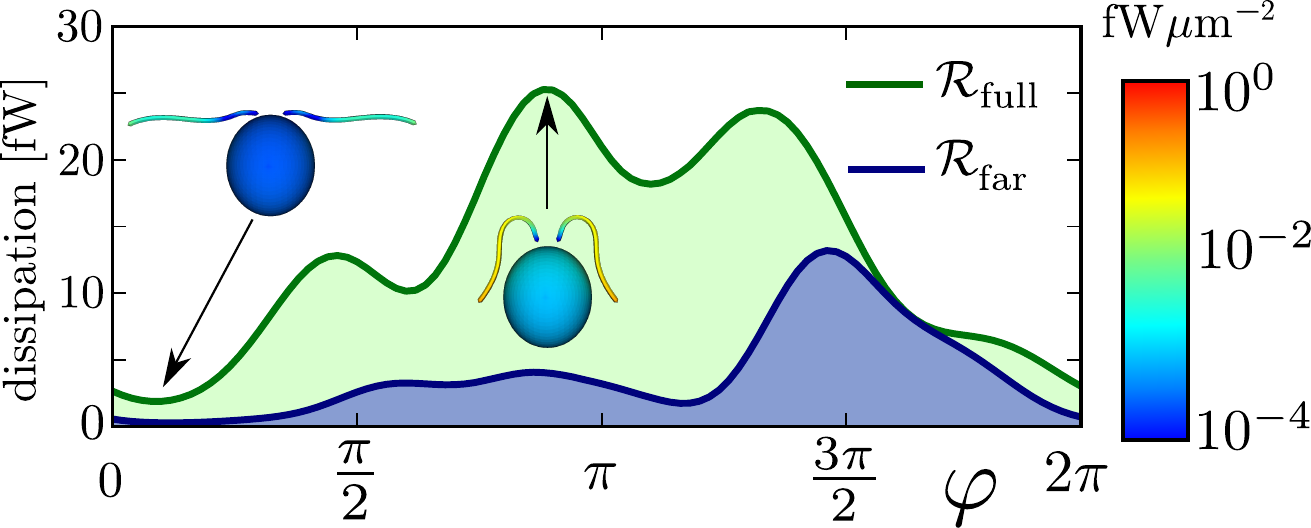}
    \caption[]{
Hydrodynamic dissipation rate varies during the beat cycle. 
We show the total rate of hydrodynamic dissipation for a swimming \textit{Chlamydomonas} cell 
as a function of flagellar phase ($\mathcal{R}_\mathrm{tot}$, red), 
and similarly the rate of dissipation in the far-field ($\mathcal{R}_\mathrm{far}$, blue), 
corresponding to its oscillating force dipole (with cut-off at $6\,\micron$). 
Insets show the surface density of work exerted by the cell on the fluid, 
which sums up to $\mathcal{R}_\mathrm{tot}$.
    }
    \label{fig:dissipation}
\end{figure}

We computed $\R(\varphi)$ for a swimming \textit{Chlamydomonas} cell, see Fig.~\ref{fig:dissipation}.
We find a pronounced phase-dependence of the rate of hydrodynamic dissipation, 
corresponding to a $45\%$ higher value for $\langle\R\rangle$ compared to a beat with reparametrized phase.
This implies that flagellar beat patterns are not optimized for minimal hydrodynamic dissipation. 
Interestingly, hydrodynamic dissipation is maximal during the flagellar recovery stroke,
for which the distance between flagella and cell body is small, 
implying high local shear rates.
Phase-dependent rates of hydrodynamic dissipation 
correspond to phase-dependent active driving forces \cite{geyer_cell-body_2013}, 
which have been proposed to facilitate hydrodynamic synchronization 
in flagellar pairs, \textit{e.g.} in \textit{Chlamydomonas} \cite{golestanian_hydrodynamic_2011}.

Previously, dissipation rates had been estimated from experimentally measured flow fields \cite{guasto_oscillatory_2010}.
These measurements represent hydrodynamic dissipation in the far field, 
as it is inherently difficult to resolve localized shear flows at distances of a few microns from the swimmer.
We therefore computed hydrodynamic dissipation in the far field, using a cut-off distance of $6\,\micron$, 
obtaining good agreement with experiment~\cite{guasto_oscillatory_2010}.

\paragraph{Conclusion.}
Using experimentally measured beat patterns, 
we computed time-dependent flow fields generated by flagellated microswimmers
and found that these oscillate between pusher- and puller-type swimming.
We presented a novel approximation scheme to account for inertial effects in
microswimming problems, and find that fluid inertia attenuates oscillatory flows 
beyond a typical distance of $100\,\micron$ from the swimmer.
Within this distance, oscillatory flows dominate any static component.
These flow signatures characterize hydrodynamic interactions between several swimmers,
which are likewise found to oscillate in time.
The static component of hydrodynamic interactions 
can be of similar magnitude as both thermal and active fluctuations of flagellar swimming.

Active swimming implies a continuous dissipation of energy into the fluid,
which we estimate as 
$300\,\mathrm{pN}\,\micron$ or $7.5\cdot 10^4\,k_B T$ per beat cycle
for a swimming \textit{Chlamydomonas} cell.
This energy equals the chemical potential of about $3\cdot 10^3$ ATP molecules \cite{jonathan_howard_mechanics_2001},
which serve as chemical fuel for the $3\cdot 10^4$ molecular motor domains that power the flagellar beat \cite{nicastro_molecular_2006}.
This implies that either only every tenth motors takes a step during each beat cycle, 
or that the efficiency of energy conversion is low.

Interestingly, 
we find that the rate of hydrodynamic dissipation 
varies during the beat cycle. 
This implies that flagellar beat patterns do not minimize hydrodynamic dissipation as sole optimization criterion. 
We speculate that design constraints of flagellar beat generation 
or internal dissipation within the flagellum might also played a role in the evolution of beat patterns. 

\textit{Acknowledgments.}
We thank V.F.~Geyer for recording flagellar beat patterns.
GSK gratefully acknowledges financial support from the DFG \textit{Microswimmers} priority program
(grant FR 3429/1-1).

\bibliography{refs} 
\bibliographystyle{steffensapsrev}

\end{document}